\documentclass{Rinton-P9x6}

\newcommand{\tr}{\mbox{tr}}
\newcommand{\zr}[1]{\mbox{\hspace*{#1em}}}
\newcommand{\ID}{\mbox{{\sf 1}\zr{-0.16}\rule{0.04em}{1.55ex}\zr{0.1}}}
\begin{document}

\title{Quantum Solitons in the Electroweak Theory}

\author{Vishesh Khemani}

\address{Center for Theoretical Physics, Laboratory for Nuclear Science and Department of Physics, Massachusetts Institute of Technology, Cambridge, MA 02139
\\E-mail: vishesh@mit.edu
}

\maketitle

\abstracts{
We explore the space of solutions of the classical equations of motion in the Euclidean electroweak theory.  We sketch a topological prescription that finds known solutions and indicates the existence of novel ones.  All spatially--varying, time--independent solutions are unstable.  However, if we consider quantum fluctuations around static classical configurations, it may be possible to find stable solutions called quantum solitons.  Such objects carry a conserved quantum number, in analogy with a topological soliton carrying a topological charge.  We explain the mechanism and motivation for the existence of a quantum soliton and describe our search for one within a spherical ansatz.  We also comment on promising candidates outside the ansatz. 
}
\section{Introduction}
The Euclidean electroweak theory has several classical solutions, like the instanton and the sphaleron,  which have rich phenomenology associated with them.  If we restrict our attention to spatially--varying, static configurations, then all known solutions are unstable and are generically called {\it sphalerons} (to distinguish them from stable solutions or {\it solitons}).  They do not have any associated quantum extended--particle states.  The discovery of a stable configuration would result in a soliton sector in the Hilbert space of states, in addition to the familiar vacuum sector.  There is no topological reason for stability of a static configuration in the electroweak theory, but a non--topological soliton (corresponding to a local minimum of the energy) may still exist.  However, in the absence of a topological beacon, it is difficult to search for such an object.  If we consider quantum fluctuations around classical configurations, then there are compelling reasons to expect the existence of {\it quantum solitons}, and well--understood mechanisms to guide the search for them.  Such objects are stabilized  by having them carry a conserved quantum number, in analogy with topological solitons carrying a topological charge.

In Sec.~\ref{Classical Sphalerons}, we briefly survey classical solutions in the bosonic sector of the Euclidean electroweak theory.  We sketch a method,  which uses non--trivial topological maps into the gauge group, to construct known solutions and propose novel ones.  In Sec.~\ref{Quantum Solitons}, we introduce the idea of quantum solitons and the mechanism and motivation for their existence.  We sketch an efficient computational method that makes it feasible to look for quantum solitons.  We describe our search within a spherical ansatz and comment on promising candidates beyond the ansatz. 
\section{Solutions in the Bosonic Sector}
\label{Classical Sphalerons}
The bosonic sector of the electroweak theory is an $SU(2) \times U(1)$ gauged Higgs theory.  For convenience and clarity we set the Abelian coupling strength to 0, which decouples that sector and allows us to ignore its dynamics.  The Lagrangian density is
\begin{equation}
\mathcal{L}_B =  -\frac{1}{2} \tr \left(F^{\mu\nu}F_{\mu\nu}\right) 
+ \frac{1}{2}\tr 
\left(\left[D^{\mu}\Phi \right]^{\dag} D_{\mu}\Phi\right) - 
\frac{\lambda}{4}\left[ 
\tr\left(\Phi^{\dag}\Phi\right) - 2v^2 \right]^2 \, ,
\label{HiggsLagrangian}
\end{equation}
where
\begin{eqnarray}
F_{\mu\nu} & = & \partial_\mu W_\nu - \partial_\nu W_\mu - i g\left[ W_\mu 
, W_\nu \right] \, , \nonumber \\
D_\mu \Phi & = & \left( \partial_\mu - i g W_\mu \right)\Phi \, , \nonumber 
\\
W_\mu & = & W_{\mu}^a \frac{\tau^a}{2} \, , 
\label{covariant}
\end{eqnarray}  
$\lambda$ is the Higgs 
self--interaction coupling constant and $v$ denotes the tree--level vacuum expectation value of the 
Higgs field. The $2 \times 2$ matrix field $\Phi$ is related to the Higgs doublet 
$\phi$ by 
\begin{equation}
\Phi =  \left( \begin{array}{cc} \phi_2^{*} & \phi_1 \\ -\phi_1^{*} & 
\phi_2 \end{array} \right) \, .
\end{equation}

There are several configurations of gauge and Higgs fields that solve the classical equations of motion.  We sketch a general prescription, which uses topologically non--trivial maps into the gauge group, to find known solutions in the Euclidean theory and motivate the existence of new ones.  The basic idea is due to Manton\cite{Manton:1983nd}, and it has been generalized by Klinkhamer\cite{Klinkhamer:2003hz}.

A finite Euclidean--action configuration is pure gauge at spacetime infinity:
\begin{equation}
W_\mu^{(\infty)} = \frac{i}{g} U \partial_\mu U^\dag \, , \Phi^{(\infty)} = v U \, , 
\label{eq:asymptotic}
\end{equation}
where $U$ is a map from the boundary of spacetime to $SU(2)$.  We allow the configuration to have trivial dimensions, in which case it has finite action per unit volume of the trivial dimensions, and the domain of $U$ is the appropriate subspace of the boundary of spacetime.  For example, a static configuration has time as a trivial dimension and in order to have finite energy (action per unit time), it must be pure gauge at spatial infinity.  Now, the third and fourth homotopy groups of $SU(2)$ are non--trivial:
\begin{equation}
\Pi_3(SU(2)) = {\mathcal Z} \, , \Pi_4(SU(2)) = {\mathcal Z}_2 \, .
\end{equation}        
So each map from $S_3$ into $SU(2)$ belongs to a homotopic class labeled by an integer and it cannot be continuously deformed into any map in a distinct class.  Similarly, each map from $S_4$ into $SU(2)$ belongs either to the trivial class (which contains the trivial map) or the non--trivial class.  Consider any topologically non--trivial map into $SU(2)$.  Identify a subspace of the domain with the boundary of spacetime spanned by the non--trivial dimensions.  Any remaining coordinates in the domain are interpolation parameters that define a sequence of configurations.  The sequence becomes a loop when we restrict all configurations on the boundary of the interpolation space to be the trivial configuration ($W_\mu=0, \Phi=v\ID$).  The non--trivial topology prohibits the loop from shrinking to a point.  The top of the tightest non--contractible loop should be an unstable solution, which is generically referred to as a sphaleron.  

Here is an example of the above construction.  Consider a winding 1 map from $S_3$ (parametrized by the angles $\beta_1, \beta_2, \alpha$) to $SU(2)$:
\begin{eqnarray}
U^{(1)}(\beta_1, \beta_2, \alpha) & = & e^{i\beta_1\tau^3}\left[ \cos(\beta_1)\ID + i \sin(\beta_1)\cos(\beta_2)\tau^3 + \right. \nonumber \\
 & & \left. i\sin(\beta_1)\sin(\beta_2) \{\cos(\alpha) \tau^1 + \sin(\alpha) \tau^2\} \right] \,
\label{eq:windingOneMap}
\end{eqnarray}
where $\beta_i \in (0,\pi)$ and $\alpha\in(0,2\pi)$.  If we identify an $S_2$ subspace of the domain with the boundary of space, we get a sequence of maps from the boundary of space to $SU(2)$:
\begin{equation}
U_{\beta_1} (\theta, \phi) = U^{(1)}(\beta_1, \theta, \phi) \, , 
\end{equation}
where $\theta, \phi$ span the spatial boundary and the remaining coordinate, $\beta_1$, is the interpolation parameter.  $U_{\beta_1}$ defines a sequence of asymptotic configurations (using Eq.~\ref{eq:asymptotic}), which is smoothly continued into the bulk of space for each $\beta_1$.  The configurations at the end points of the sequence ($\beta_1=0,\pi$) are chosen to be trivial (which is possible because $U_{\beta_1}$ is the identity at the end points) and we have a loop of configurations.  Suppose we could continuously deform the loop so that for every $\beta_1$ the configuration is trivial.  Then $U^{(1)}(\beta_1, \theta, \phi)$ could be continuously deformed into the trivial map, which is impossible.  So the loop is non--contractible and this indicates the existence of an unstable, static solution at the top of the tightest loop.  This can be easily found by a straightforward continuation into all space of the asymptotic configuration at $\beta_1=\pi/2$.  It is the well--known {\it weak sphaleron}\cite{Manton:1983nd,Klinkhamer:1984di}.  It is the lowest barrier between topologically inequivalent vacua and its energy determines the rate of fermion number violating processes at temperatures comparable to the electroweak phase transition scale. 

If we identify the whole domain of the map in Eq.~\ref{eq:windingOneMap} with the boundary of spacetime then there are no remaining interpolation parameters.  Instead there is a topologically stable solution -- the {\it weak instanton}\cite{Belavin:fg}.  It describes fermion number violation via tunneling.  If we consider static configurations with one trivial dimension (say $z$) and identify an $S_1$ subspace of the $S_3$ domain with the boundary of the $x-y$ plane, then we get a two--parameter non--contractible loop of configurations, the top of which is the {\it W--string}\cite{Vachaspati:fi,Klinkhamer:1994uy}.  Its instability detracts from its significance, especially with regard to electroweak baryogenesis where it could have played a crucial role (however, see Sec.~\ref{BeyondTheSpherical}).

The above procedure using a winding $n$ map from $S_3$ to $SU(2)$ indicate the existence of {\it multi--instantons}, {\it multi--sphalerons}\cite{Kleihaus:1994yj} and {\it winding $n$ W--strings}.  Finally, if we use a non--trivial map from $S_4$ to $SU(2)$ to construct non--contractible loops, there is evidence for novel solutions: an {\it $I^*$} (a topologically trivial Euclidean solution with one unstable direction)\cite{Klinkhamer:1996ad}, an {\it $S^*$} (a static solution with two unstable directions)\cite{Klinkhamer:1990ik} and a {\it $W^*$} (a static solution with one trivial dimension and three unstable directions).  We are currently investigating the existence and significance of these.

Note that these topological arguments do not guarantee the existence of the solutions described above, because the configuration space is a non--compact manifold and the non--contractible loops may run off to infinity.  Nor is it clear that two different loops give two distinct solutions.  Nevertheless, the topology points to possible solutions in the vast configuration space and once we know where to look, we can verify whether a solution exists.

We find that all known (and proposed) time--independent solutions are unstable.  The classical bosonic sector seems to have only sphalerons and no solitons.  Of course, the existence of non--topological solitons (corresponding to local minima of the energy functional) cannot be excluded.  However, having exhausted the topological properties of the theory, we are left with no guiding principle to enable a search for such objects.  But if we consider quantum effects on the classical bosonic sector, then there are compelling reasons to expect the existence of quantum solitons and well--understood mechanisms to guide the search for them.  
\section{Quantum Solitons}
\label{Quantum Solitons}
\subsection{The Idea}
\label{Idea}
A topological soliton is a non-vacuum, static configuration that is topologically stable.  It carries a conserved topological charge which prevents it from decaying into a vacuum configuration with no topological charge.  Analogously, it may be possible to stabilize a configuration by making it carry a conserved quantum number (say fermion number).  We use the term quantum soliton to refer to any such quantum--stabilized object.

There is a natural mechanism in the electroweak theory for the existence of quantum solitons.  There are several known configurations in the bosonic sector that tightly bind fermions (quarks and leptons) in their vicinity.  These configurations consist of classical solutions (the sphalerons discussed in Sec. \ref{Classical Sphalerons}) as well as non-solutions.  The existence of tightly--bound levels suggests that it may be energetically favorable for a certain number of fermions (say $N_f$) to be trapped by such backgrounds, with a small associated occupation energy, $E_{\rm occ}^{(N_f)}$.  The binding energy could outweigh the cost in classical energy, $E_{\rm cl}$, to set up the configuration, i.e. $E_{\rm cl}+E_{\rm occ}^{(N_f)} < m_f N_f$ where $m_f$ is the mass of each perturbative fermion.  However, to be consistent to order $\hbar$, we must also include the Casimir energy: the renormalized energy shifts of all the other fermion modes.  For static configurations this is the renormalized one--loop fermion vacuum energy $E^{\rm ren}_{\rm vac}$.  Thus, the minimum total energy of $N_f$ fermions associated with an arbitrary configuration $C=\{W, \Phi\}$ is
\begin{equation}
E^{(N_f)}[C] = E_{\rm cl}[C] + E_{\rm occ}^{(N_f)}[C] + E_{\rm vac}^{\rm ren}[C] \, .
\label{ENf}
\end{equation}
If $C$ decays into a vacuum configuration then it must create $N_f$ perturbative fermions (ignoring anomalous violation of fermion number, which is exponentially suppressed).  However, the strong fermion binding suggests that there exist configurations such that $E^{(N_f)}[C]<m_f N_f$.   Then, $C$ can decay only into a quantum soliton at which $E^{(N_f)}$ has a local minimum.

The existence of such fermionic quantum solitons would provide an attractive resolution to the decoupling puzzle in the standard model (and other chiral gauge theories).  A fermion obtains its mass through Yukawa coupling to the scalar Higgs via the well-known Higgs mechanism.  Explicit mass terms are prohibited by gauge--invariance.  So, as we increase the mass of a fermion (thereby making the denominator in the propagator suppress loop corrections) we also increase the Yukawa coupling which gives a corresponding enhancement from the vertices.  Moreover, the heavy fermion cannot simply disappear from the spectrum because then anomaly cancellation would be ruined.  However, it is plausible that the large Yukawa coupling gives rise to a quantum soliton in the low energy theory.  This carries the quantum numbers of the decoupled fermion and maintains anomaly cancellation using the mechanism described by D'Hoker and Farhi\cite{D'Hoker:1983kr}.
\subsection{The Computational Method}
\label{ComputationalMethod}
The search for a quantum soliton with fermion number $N_f$ requires an exploration of $E^{(N_f)}[W,\Phi]$ defined in Eq.~(\ref{ENf}).  The renormalized one--loop fermion vacuum energy, $E_{\rm vac}^{\rm ren}$, is the only computationally intensive component of $E^{(N_f)}$.  Since we are interested in heavy fermions with large Yukawa couplings, and configurations with typical widths of the order of $1/m_f$, we cannot accurately evaluate $E_{\rm vac}^{\rm ren}$ perturbatively (using an expansion in couplings or derivatives).  Instead we use methods involving scattering data of fermions to compute it exactly and efficiently\cite{PhaseshiftsGeneral,PhaseshiftsFermions,Leipzig}.  

The unrenormalized fermion vacuum energy is given by a sum over the shift in the zero-point energies of the fermion modes due to the background bosonic configuration.  We write this as a sum over bound state energies, $\epsilon_j$, and a momentum integral over the continuum state energies weighted by the change in the density of states, $\Delta \rho(k)$,
\begin{equation}
E_{\rm vac} = -\frac{1}{2}\sum_j|\epsilon_j| - 
\frac{1}{2}\int_0^\infty dk \sqrt{k^2+m_f^2}\, \Delta\rho(k) \, .
\end{equation}     
The above integral is rendered finite by subtracting a sufficient number of terms (say $N$) in the Born series expansion of $\Delta \rho$ and adding back in exactly the same (divergent) quantity in the form of one--fermion--loop Feynman diagram contributions with $i$ insertions of the background potential, $E_{\rm FD}^{(i)}$:
\begin{eqnarray}
E_{\rm vac} & = & -\frac{1}{2}\sum_j |\epsilon_j| 
- \frac{1}{2}\int_0^\infty dk \sqrt{k^2+m_f^2} 
\left( \Delta\rho(k) - \sum_{i=1}^{N}\Delta\rho^{(i)}(k)\right) 
\nonumber\\ & &  
+ \sum_{i=1}^N E_{\rm FD}^{(i)} \, .
\end{eqnarray}
Now all divergences are isolated in low--order Feynman diagrams and we perform conventional perturbative renormalization in quantum field theory.  We decompose the bare parameters in the bosonic sector, defined by the Lagrangian in Eq.~(\ref{HiggsLagrangian}), into renormalized parameters and counterterm coefficients, and get a (divergent) counterterm contribution, $E_{\rm ct}$, to the energy.  We fix the regulated counterterm coefficients once and for all by specifying physical parameters such as the Higgs vacuum expectation value and the masses of the Higgs and the gauge bosons, independent of the background configuration.  We add the counterterm contribution to the regulated Feynman diagrams contribution and remove the regulator to obtain the finite, renormalized energy,
\begin{equation}
E_{\rm vac}^{\rm ren} = E_{\rm vac} + E_{\rm ct} \, .
\end{equation}  

The change in the density of states, $\Delta \rho(k)$, is obtained from the momentum derivative of the phase shifts, induced by the background fields, of the fermion scattering wave--functions, 
\begin{equation}
\Delta\rho(k) = \frac{1}{\pi}\frac{d}{dk}\sum_G D_G \delta_{G}(k) \, .
\end{equation}
Here we have decomposed the scattering problem into partial waves labeled by $G$ (with a degeneracy $D_G$), by restricting the configurations to be sufficiently symmetric.  For any such background, the Dirac equation may be easily solved numerically for all $G$ to obtain the bound states and the scattering phase shifts (together with their Born series) required for the computation of $E_{\rm vac}$.

Thus, we have an efficient way to numerically compute the fermion one--loop vacuum energy non--perturbatively, with conventional quantum field theory renormalization in the perturbative sector of the theory.  This makes it feasible to search for quantum solitons, which are local minima of $E^{(N_f)}$.
\subsection{The Spherical Ansatz}
In this section we briefly describe our search for a quantum soliton in the electroweak theory, within a spherical ansatz (see Ref.\cite{SphericalAnsatz} for details).  The bosonic configurations are chosen to be invariant under simultaneous rotations in physical space and isospin space\cite{RatraYaffe}.  This allows a partial wave decomposition labeled by the grand spin (vector sum of angular momentum and isospin) quantum number $G$.

We carry out a variational search, looking for a configuration $C$ such that $E^{(1)}[C]<m_f$ and
$E^{(1)}[C]<E_{\rm q.s.}$, where $E_{\rm q.s.}$ is the energy of the quantum-corrected weak sphaleron.  The first condition ensures that $C$ cannot simply decay into a vacuum configuration plus a perturbative fermion.  The second condition ensures the exponential suppression of fermion number violating processes.  In other words, it prevents $C$ from rolling over the weak sphaleron, giving up its fermion number and then rolling down to a vacuum configuration.  Finding a configuration with these
properties would guarantee the existence of a nontrivial local minimum of $E^{(1)}$, i.e. a fermionic quantum soliton.  

In the vast configuration space of gauge and Higgs fields, we restrict our exploration to perturbations around backgrounds that tightly bind fermions.  These make promising candidates for a quantum soliton as explained in Sec. \ref{Idea}.  In the spherical ansatz we know of two such configurations: twisted Higgs and the weak sphaleron.

\begin{figure}[ht]
\begin{center}
\epsfxsize=20pc 
\epsfbox{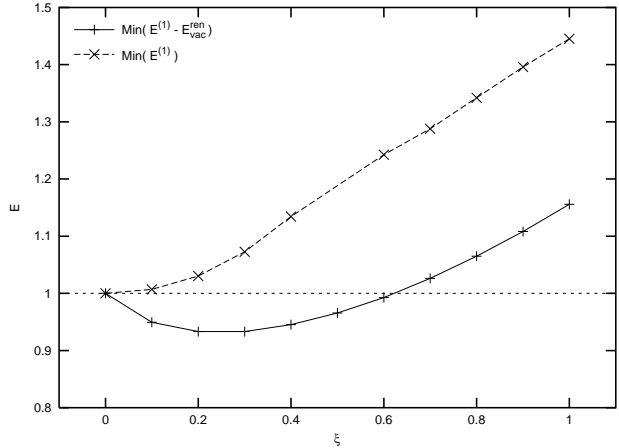} 
\caption{Minimum one--fermion energies (in units of $m_f$), with and without $E_{\rm vac}^{\rm ren}$ contributions, along the interpolation in Eq.~\ref{eq:twistedHiggsInterpolation}.}
\label{fig:destabilized}
\end{center}
\end{figure}

A twisted Higgs configuration has trivial gauge fields and scalar fields of the form
\begin{equation}
\Phi = v U^{(1)} \, 
\end{equation}
where $U^{(1)}$ is a winding 1 map from compactified space to $SU(2)$.  It is not a classical solution but nevertheless strongly binds a fermion.  We smoothly interpolate from the trivial configuration to a twisted Higgs (which is possible as long as $\Phi^\dag \Phi$ vanishes at some point in space along the interpolation), using an interpolating parameter $\xi$, which goes from 0 to 1:
\begin{equation}
\Phi = v(1-\xi)\ID + v\xi \mbox{exp} \left( -i\pi e^{-r/w} \vec{\tau}\cdot\hat{x} \right) \, .
\label{eq:twistedHiggsInterpolation} 
\end{equation}
Here $w$ is a variational parameter that characterizes the configuration width.  The lowering of the occupation energy, $E_{\rm occ}^{(1)}$, as the twisted Higgs is approached is offset by the rising classical energy, $E_{\rm cl}$, and the fermion vacuum energy, $E_{\rm vac}^{\rm ren}$.  We must investigate whether the gain in binding energy is washed out or not.  We choose a Yukawa coupling of 10 and a Higgs mass of $v/\sqrt{2}$.  For each value of $\xi$ we minimize $E^{(1)}-E^{\rm ren}_{\rm vac} = E_{\rm cl}+E^{(1)}_{\rm occ}$ and $E^{(1)}$ with respect to $w$ and plot the results in Fig. \ref{fig:destabilized}.  If $E_{\rm vac}^{\rm ren}$ is neglected, then we have configurations with energies lower than the perturbative fermion for $\xi<0.6$, indicating the existence of a soliton.  However, the $E_{\rm vac}^{\rm ren}$ contribution raises the energies to above $m_f$.  We find that in general, the fermion vacuum energy destabilizes would--be solitons, such as those found in previous work\cite{NolteKunz}.  We find no evidence for a fermionic soliton, even when we consider various other perturbations around a twisted Higgs.  

\begin{figure}[t]
\begin{center}
\epsfxsize=20pc 
\epsfbox{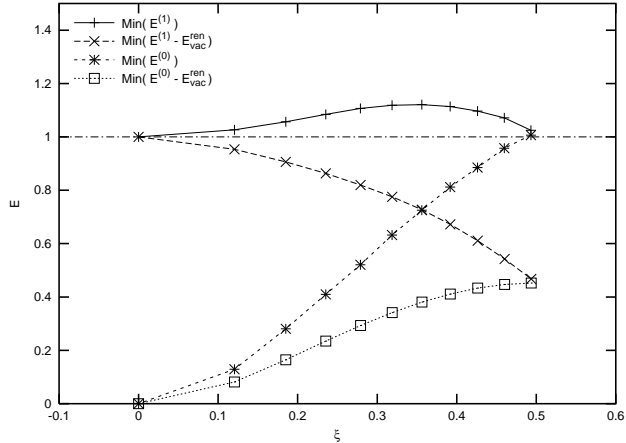} 
\caption{Minimum energies (in units of $m_f$) along the interpolation in Eq.~\ref{eq:sphaleronInterpolation} in both the zero--fermion and one--fermion sectors (with and without $E_{\rm vac}^{\rm ren}$ contributions).}
\label{fig:sphaleron}
\end{center}
\end{figure}

Next we consider paths from the trivial vacuum configuration (with winding 0) to a winding 1 vacuum configuration.  Along such paths, a fermion level leaves the positive continuum, crosses zero at the sphaleron, and finally enters the negative continuum.  Again the question is whether the gain in binding energy, as we approach the sphaleron along the interpolation, is sufficient to stabilize a configuration.  Consider the linear interpolation given by
\begin{equation}
\Phi = v(1-\xi)\ID + v \xi U^{(1)} \, , W_j = \xi \frac{i}{g}U^{(1)}\partial_j U^{(1)\dag} \, , 
\label{eq:sphaleronInterpolation}
\end{equation}
where 
\begin{equation}
U^{(1)} = \mbox{exp}\left( -i\pi e^{-r/w} \vec{\tau}\cdot\hat{x} \right) \, ,
\end{equation}
and the configuration width $w$ is a variational parameter.  We choose the following theory parameters: Yukawa coupling of 10, Higgs mass of $v/\sqrt{2}$ and gauge coupling of 6.5.  In Fig.~\ref{fig:sphaleron} we plot zero--fermion and one--fermion energies, minimized within the above variational ansatz, for different values of $\xi$, both with and without the $E_{\rm vac}^{\rm ren}$ contributions.  First consider the zero--fermion case with the two curves $E^{(0)}-E_{\rm vac}^{\rm ren}=E_{\rm cl}$ and $E^{(0)}$.  At $\xi=1/2$, the classical sphaleron is the configuration at which $E_{\rm cl}$ is minimum, and a distinct configuration minimizes $E_{\rm cl}+E_{\rm vac}^{\rm ren}$.  This is the quantum--corrected sphaleron, which has an energy more than double that of the classical sphaleron (for the theory parameters considered).  Next consider the data in the one--fermion sector for $E^{(1)}-E_{\rm vac}^{\rm ren}=E_{\rm cl}+E^{(1)}_{\rm occ}$ and $E^{(1)}$, also shown in Fig.~\ref{fig:sphaleron}.  Since the
classical sphaleron has an energy much smaller than $m_f$, one would expect that the perturbative fermion would have an unsuppressed decay mode over the
sphaleron, as first pointed out by Rubakov\cite{Rubakov}. 
The $E^{(1)} -E_{\rm vac}^{\rm ren}$ curve indeed displays this decay path.  The
fermion vacuum polarization energy modifies things in two crucial
ways.  First, the fermion quantum corrections to the sphaleron raise
its energy to be degenerate with the fermion, as mentioned before.  So
the threshold mass is significantly increased.  Second, in the plot of
$E^{(1)}$ we observe that there is an energy barrier 
between the fundamental fermion
and the quantum-corrected sphaleron.  This indicates that even when
the fermion becomes heavier than the sphaleron, there might exist a
range of masses for which the decay continues to be exponentially
suppressed (since it can proceed only via tunneling).  When we consider other paths over the sphaleron (such as an instanton with Euclidean time being the interpolating parameter), we find that the energies minimized in the linear interpolation ansatz are not lowered significantly.  Thus, the fact that the minimum--$E^{(1)}$ surface does not have a minimum in this ansatz indicates the absence of a fermionic soliton.
\subsection{Beyond the Spherical Ansatz}
\label{BeyondTheSpherical}
In hindsight it is not too surprising that we do not find a quantum soliton within the spherical ansatz.  As mentioned in Sec.~\ref{Idea}, the existence of such solitons would maintain anomaly cancellation when fermions are decoupled from the electroweak theory.  Without the hypercharge gauge field, the only anomaly is Witten's global anomaly\cite{Witten} due to topologically non-trivial maps from $S_4$ to $SU(2)$.  However, in the spherical ansatz, the theory reduces to a $U(1)$ theory in which $\Pi_3(SU(2))$ persists as $\Pi_1(U(1))$ (and so we have topologically inequivalent vacua and the weak sphaleron) but there is no remnant of $\Pi_4((SU(2))$.  So, the quantum soliton that could resolve the decoupling puzzle probably lies outside the ansatz.

One route beyond the spherical ansatz is the use of non--trivial $\Pi_4(SU(2))$ to construct non--contractible loops and corresponding novel classical sphalerons (see Sec.~\ref{Classical Sphalerons}).  We expect fermion zero--modes in these sphaleron backgrounds and thus the required tight--binding--mechanism exists.  Moreover, since the non--trivial topology that gives rise to Witten's anomaly is built into the construction of such configurations, they are promising candidates for objects that maintain anomaly cancellation in the decoupled theory.  We are currently exploring this possibility.
  
Another interesting non--spherical configuration is the W--string solution mentioned in Sec.~\ref{Classical Sphalerons}.  Our motivation goes beyond fermion decoupling.  Stable electroweak strings are a crucial ingredient in a scenario for electroweak baryogenesis even without a first order electroweak phase transition\cite{Brandenberger}.  It is likely that the classically unstable W--string could be quantum--stabilized by having it carry fermion number.  The string configuration gives rise to fermion zero modes and it is possible that that several massless quarks and leptons are trapped along the string in a stable manner.  We are presently investigating this.
\section{Conclusions}
The bosonic sector of the electroweak theory has several classical solutions, which we can find using topologically non--trivial maps into the gauge group.  All spatially--varying, static solutions turn out to be unstable.  However, when we consider quantum--fluctuations around classical configurations, then it is likely that there exist quantum solitons that carry some fermion number.  We have an efficient way of computing the energy of fermionic configurations, thus making a search for a quantum soliton feasible.  We do not find such an object within a spherical ansatz.  Nevertheless there are promising non--spherical candidates: novel classical sphalerons and electroweak strings.  We continue to explore these possibilities.
\section*{Acknowledgments}
This work has been done in collaboration with E.~Farhi, N.~Graham, R.~L.~Jaffe, M.~Quandt, O.~Schroeder and H.~Weigel.  I am grateful to the organizing committee of QFEXT03 for providing support for my participation in the workshop.  I am supported in part by the U.S. Department of Energy (D.O.E.) under cooperative research agreement \#DF-FC02-94ER40818.


\newpage
\begin{thebibliography}{99}
\bibitem{Manton:1983nd}
N.~S.~Manton,
Phys.\ Rev.\ D {\bf 28}, 2019 (1983).
%
\bibitem{Klinkhamer:2003hz}
F.~R.~Klinkhamer and C.~Rupp,
J.\ Math.\ Phys.\  {\bf 44}, 3619 (2003)
[arXiv:hep-th/0304167].
%
\bibitem{Belavin:fg}
A.~A.~Belavin, A.~M.~Polyakov, A.~S.~Shvarts and Y.~S.~Tyupkin,
Phys.\ Lett.\ B {\bf 59}, 85 (1975).
%
\bibitem{Klinkhamer:1984di}
F.~R.~Klinkhamer and N.~S.~Manton,
Phys.\ Rev.\ D {\bf 30}, 2212 (1984).
%
\bibitem{Vachaspati:fi}
T.~Vachaspati,
Phys.\ Rev.\ Lett.\  {\bf 68}, 1977 (1992)
[Erratum-ibid.\  {\bf 69}, 216 (1992)].
%
\bibitem{Klinkhamer:1994uy}
F.~R.~Klinkhamer and P.~Olesen,
Nucl.\ Phys.\ B {\bf 422}, 227 (1994)
[arXiv:hep-ph/9402207].
%
\bibitem{Kleihaus:1994yj}
B.~Kleihaus and J.~Kunz,
Phys.\ Lett.\ B {\bf 329}, 61 (1994)
[arXiv:hep-ph/9403289].
%
\bibitem{Klinkhamer:1996ad}
F.~R.~Klinkhamer and J.~Weller,
Nucl.\ Phys.\ B {\bf 481}, 403 (1996)
[arXiv:hep-ph/9606481].
%
\bibitem{Klinkhamer:1990ik}
F.~R.~Klinkhamer,
Phys.\ Lett.\ B {\bf 246}, 131 (1990).
%
\bibitem{D'Hoker:1983kr}
E.~D'Hoker and E.~Farhi,
Phys.\ Lett.\ B {\bf 134}, 86 (1984).

%
\bibitem{PhaseshiftsGeneral}
E.~Farhi, N.~Graham, P.~Haagensen and R.~L.~Jaffe,
Phys.\ Lett.\ B {\bf 427}, 334 (1998)
[arXiv:hep-th/9802015].
\bibitem{PhaseshiftsFermions}
N.~Graham and R.~L.~Jaffe,
Nucl.\ Phys.\ B {\bf 549}, 516 (1999)
[arXiv:hep-th/9901023].
%
\bibitem{Leipzig}
N.~Graham, R.~L.~Jaffe and H.~Weigel,
Int.\ J.\ Mod.\ Phys.\ A {\bf 17}, 846 (2002)
[arXiv:hep-th/0201148].
%
\bibitem{SphericalAnsatz}
E.~Farhi, N.~Graham, R.~L.~Jaffe, V.~Khemani and H.~Weigel,
Nucl.\ Phys.\ B {\bf 665}, 623 (2003)
[arXiv:hep-th/0303159].
%
\bibitem{RatraYaffe}
B.~Ratra and L.~G.~Yaffe,
Phys.\ Lett.\ B {\bf 205}, 57 (1988).
%
\bibitem{NolteKunz}
G.~Nolte and J.~Kunz,
Phys.\ Rev.\ D {\bf 48}, 5905 (1993)
[arXiv:hep-ph/9308290].
%
\bibitem{Rubakov}
V.~Rubakov,
Nucl.\ Phys.\ B {\bf 256}, 509 (1985).
%
\bibitem{Witten}
E.~Witten
Phys.\ Lett.\ B {\bf 117}, 324 (1982).
%
\bibitem{Brandenberger}
R.~H.~Brandenberger and A.~C.~Davis,
Phys.\ Lett.\ B {\bf 308}, 79 (1993)
[arXiv:astro-ph/9206001].
%
\end{thebibliography}
\end{document}